\documentclass[letter]{aa}  

\usepackage{graphicx}
\usepackage{txfonts}
\usepackage{amssymb}

\newcommand{\kms}{${\rm km\ s^{-1}}$}
\newcommand{\dunit}{${\rm cm^2\,s^{-1}}$}
\newcommand{\pcc}{${\rm cm^{-3}}$}

\newcommand{\sect}{section}
\newcommand{\change}[1]{{#1}} 
\newcommand{\chtwo}[1]{{#1}} 

\begin{document}

   \title{Superbubbles as Galactic PeVatrons: The Potential Role of Rapid Second-Order Fermi Acceleration}


   \author{Jacco Vink
          \inst{1,2}
          }

   \institute{
     Anton Pannekoek Institute/GRAPPA, University of Amsterdam, Science Park 904, 1098 XH Amsterdam, The Netherlands\\
     \email{j.vink@uva.nl}
     \and
     SRON Netherlands Institute for Space Research, Niels Bohrweg 4, 2333 CA Leiden, The Netherlands
   }

   \date{Received ;accepted }

 
  \abstract
   {The origin of Galactic cosmic rays is still a mystery, in particular the sources and acceleration mechanism for cosmic rays with energies up to or beyond a PeV. Recently LHAASO has and H.E.S.S have shown that two gamma-ray sources associated with superbubbles created by young massive stellar clusters are likely PeVatrons. This has renewed the interest in the cosmic-ray acceleration processes in superbubbles.
   }
   {To study the possibility and conditions under which  second-order Fermi acceleration can accelerate particles beyond PeV energies in superbubbles.}
   {An analytical equation is derived for the maximum  energy a cosmic-ray particle can obtain as a function of acceleration duration \change{and size}. The maximum energy depends critically on the diffusion coefficient $D$ and the Alfv\'en velocity, $V_{\rm A}$.   The analytical solutions for the acceleration  time scale shows that second-order Fermi acceleration can be just as efficient as diffusive shock acceleration, when comparable relevant velocities are used---i.e.  the Alfv\'en velocity or shock velocity. \change{Ultimately, the maximum energy is determined by the diffusive escape of the cosmic rays, which depends on the size and diffusion coefficient.
   This limits the maximum energy to a few PeV for superbubble radii of $\sim 50$~pc.}
   The probable values for the diffusion coefficient and Alv\'en speed are  studied for two likely PeVatron regions, HESS J1646-458 associated with Westerlund 1, and the Cygnus Cocoon, associated Cyg OB2.
   }
   {
   It is shown that within a typical stellar cluster time scale of 1-5~Myr cosmic rays can be accelerated to $>10^{15}$~eV, provided that  $V_{\rm A} \gtrsim 300$~\kms, and the diffusion coefficient is
   $D\sim 10^{26}$~\dunit at 100~TeV.    This suggests that second-order Fermi acceleration in superbubbles should be considered as a possible source of Galactic cosmic rays up to, or beyond a PeV.
     }
   {}


   \maketitle
%

   \nolinenumbers

\section{Introduction}

The origin of  Galactic cosmic rays \change{(CRs)}---with energies of $\sim 0.1$~GeV to $\sim 3 \times 10^{15}$~eV or even  $\sim 3 \times 10^{18}$~eV---is
a longstanding problem. The energy budget to maintain the energy density of CRs in the Galaxy is fulfilled by supernovae, and young supernova remnants (SNRs) have indeed, through X-ray and gamma-ry observations,
 been well-established as important sources of CRs, but only up to $\sim$10~TeV--100~TeV \citep[][for a review]{helder12a}. However, it is very questionable whether
  young SNRs can regularly accelerate CRs up to,
 or even beyond, 3~PeV \citep[e.g.][]{cristofari21}, the energy of the break---the ``knee" ---in the CR spectrum.
 This problem of the origin of the highest energy Galactic CRs
  has been popularized as a quest for Galactic CR PeVatrons \citep[][]{gabici07}.

\change{An incomplete} answer has been provided by TeV to PeV gamma-ray observations. LHAASO has detected 100-1400~TeV photons providing localizations of PeVatrons \change{\citep{lhaaso21,lhaaso24}.} But 
at lower  energies  also  experiments like
HAWC \citep{hawc20,hawc22_gro} and atmospheric Cherenkov telescopes like H.E.S.S. \citep[e.g.][]{hess21_1702,hess22_wd1}, and MAGIC \citep{magic21_g109} 
contributed to localizing  \change{PeVatron candidates}.
However, pinpointing the exact acceleration locations of  PeV CRs proves to be difficult, except for a case like the Crab Nebula,
which is likely an accelerator of electrons/positrons (leptons), which cannot explain the origin of  the hadronic CRs that dominate the CR spectrum.  
Instead, many PeVatrons identified so far are associated with starforming regions, but these contain multiple energetic sources, such as SNRs and pulsar wind nebulae,
and it is not clear which of these, if any in particular, is the main source of PeV CRs.
Nevertheless, one type of PeV sources stand out as potential hadronic PeVatrons: the superbubbles \change{(SBs)} created by young massive  stellar clusters.  
Two prominent candidates are  LHAASO J2032+4102 \change{\citep{lhaaso24_cygnus}},  associated with the
Cygnus Cocoon/Cyg OB2 \citep{fermi11_cocoon,hawc21_cygnus}, and  HESS J1646-458  \citep{hess22_wd1} associated with Westerlund 1.

The idea that massive young stellar clusters/OB associations are the likely locations of powerful CR accelerations, through the combined effects of shocks induced by strong stellar winds and supernovae
has been around for a few decades \citep{montmerle79,bykov92,parizot04}. The rationale is that the localized combination of powerful sources of energies result in collective effects, such as colliding shocks, strong
plasma turbulence, and prolonged acceleration times, that provide more efficient CR acceleration than is possible by the individual sources that make up the cluster.
Recently there have been several studies that focus on the role of the collective wind termination shock as the prime source of CR acceleration in SBs due to
diffusive shock acceleration (DSA) \citep{morlino21,vieu22,blasi23}. \change{In addition, some attention has been focused on the role of SNRs inside SBs for CR acceleration \citep{vieu23}, 
as they have different acceleration properties than ``field" SNRs, due to the tenuous and turbulent medium inside SBs \citep[e.g.][]{hess_lmc,kavanagh19}.}

Here I will discuss that SBs are also good locations for second-order Fermi acceleration up to PeV energies, provided the Alfv\'en velocity is large and magnetic-field turbulence is high, two conditions
that are likely realized in SBs. 


\section{Second-order Fermi acceleration}

Second-order Fermi acceleration \citep{fermi49} is caused by charged CR particles interacting with turbulent magnetic fields, which act as particle scattering centers and which move 
typically with the Alfv\'en velocity \change{
\begin{equation}\label{eq:v_a}
V_{\rm A}=\frac{B}{\sqrt{4\pi \rho}}=18.5\left(\frac{B}{10~{\rm \mu G}}\right)\left(\frac{n_{\rm H}}{1~{\rm cm^{-3}}}\right)^{-1/2}{\rm km\ s^{-1}}.
\end{equation}
}
Each elastic scattering causes a relative increase, or a decrease, in momentum due a Lorentz boost, depending on whether the scattering 
was a head-on, or a head-tail collision. 
On average the particle will gain a fractional energy of 
\begin{equation}\label{eq:dE_E}
\frac{\Delta E}{E} \approx \xi \left(\frac{v}{c}\right)^2
\end{equation}
per scattering, 
with  $c$ the particle velocity, $v\approx v_{\rm A}$. the velocity of the scattering centers, and $\xi$ a factor of order one. 

Scattering on  magnetic irregularities is also responsible for diffusion in space.  
It is common to model the Galactic diffusion of CRs using a spatial
diffusion coefficient parameterized as $D=D_0 (E/E_0)^\delta$ with $\delta\approx 0.3-0.7$ \citep[e.g.][]{strong07}.
\change{
Since 
\begin{equation}\label{eq:diff}
D=\frac{1}{3}\lambda_{\rm mfp} c = \frac{1}{3} \eta r_{\rm g} c =\frac{1}{3}\eta\frac{cE}{eZB},
\end{equation} 
with  $\lambda_{\rm mfp}$ the typical particle mean free path, and $\lambda_{\rm mfp}=\eta r_{\rm g}$ the
mean free path expressed in factors $\eta$ of the gyroradius. The smallest diffusion coefficients assume
$\eta=1$ (Bohm diffusion) and if applicable for a large range of energies, implies $\delta=1$.
From the above the implied,} typical, collision time can be expressed as  $\Delta t=\lambda_{\rm mfp}/( c)=3D/c^2$.

Assuming that the time scale for fractional energy increase is roughly equal to pitch-angle scattering, 
 we can convert (\ref{eq:dE_E}) in an energy gain rate, 
\begin{equation}\label{eq:2fermi}
\frac{1}{E}\frac{dE}{dt} \approx \frac{1}{E} \frac{\Delta E}{\Delta t}=
  \xi \frac{1}{3D_0}\left(\frac{E}{E_0}\right)^{-\delta} V_{\rm A}^2, 
\end{equation}
which has the following solution for the maximum energy, assuming an injection energy of $E_{\rm inj}$:
\begin{equation}\label{eq:emax}
E_{\rm max}=
\left[
E_{\rm inj}^\delta
+
 \frac{\delta \xi }{3D_0}  V_{\rm A}^2E_0^\delta t
\right]^{1/\delta}.
\end{equation}
For $E_{\rm max}\gg E_{\rm inj}$ this implies an acceleration time of 
\begin{equation}\label{eq:tau_2}
\tau_{\rm acc,2nd}\approx \frac{3D_0}{\delta \xi V_{\rm A}^2}\left(\frac{E_{\rm max}}{E_0}  \right)^\delta =  \frac{3D(E_{\rm max})}{\delta \xi V_{\rm A}^2}.
\end{equation}

\begin{figure}

  \centerline{
    \includegraphics[trim=0 0 0 0,clip=true,width=0.99\columnwidth]{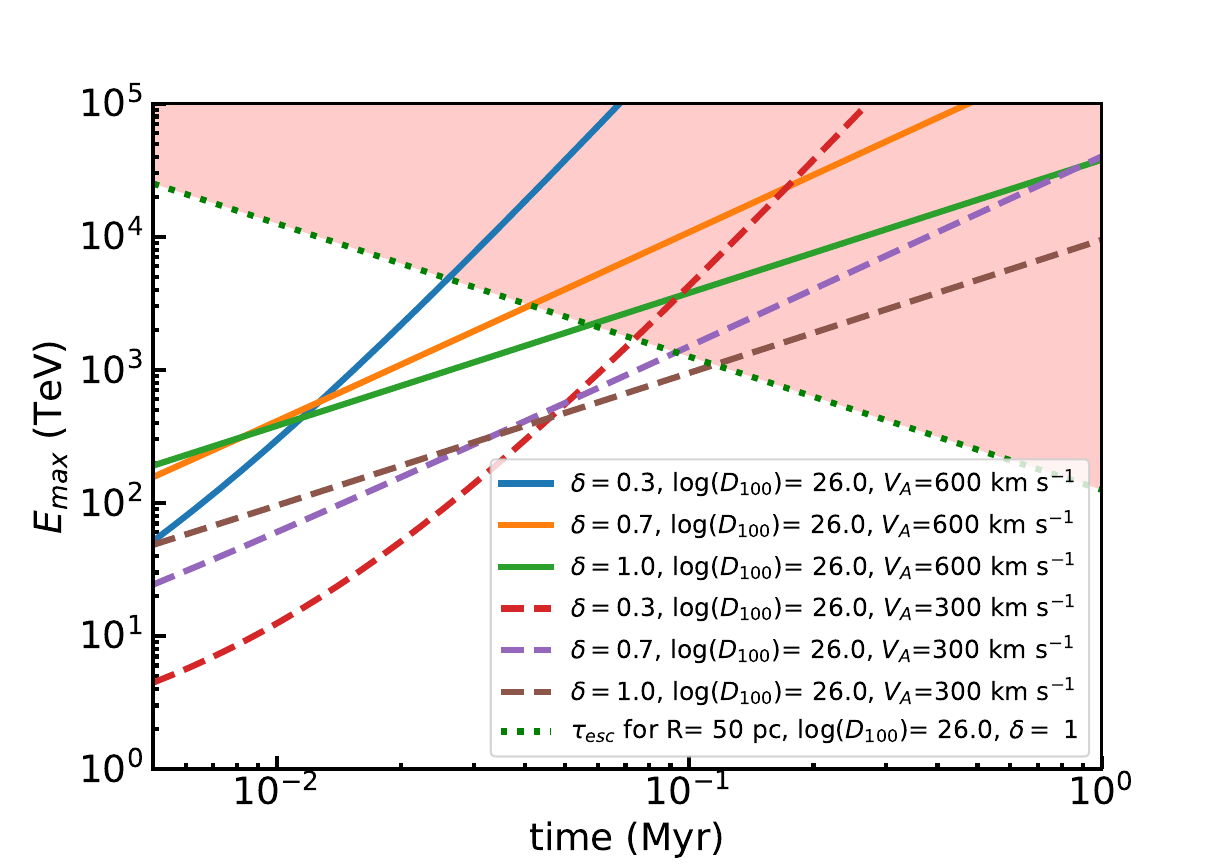}
}
  \caption{
    \label{fig:emax}
 \change{   The maximum energy as a function of time according to (\ref{eq:emax})
    for various combinations of the Alfv\'en velocity and $\delta$, and
    an assumed injection energy of 1~TeV, using $\xi=1$. $D=2\times 10^{26}$~\dunit\ at 100~TeV.
    The dotted line indicates the escape time $\tau_{\rm esc}$ (Eq.~\ref{eq:tau_esc}) for $\delta=1$ (Bohm diffusion) and $R=50$~pc.
    The shaded area is excluded based on escape time constraints.
    See \sect~\ref{sec:ymc} for the choice of parameter values.}
  }
\end{figure}

We can compare this to the acceleration time scale for DSA \citep[see Eqs.~(11.37) \& (11.38) in][]{vinkbook}:
\begin{equation}\label{eq:tau_1}
\tau_{\rm acc,1st}\approx 
\frac{8 D_0}{\delta V_{\rm s}^2}\left(\frac{E_{\rm max}}{E_0}\right)^\delta =\frac{8 D(E_{\rm emax})}{\delta V_{\rm s}^2},
\end{equation}
with $V_{\rm s}$ the shock velocity, and using  a shock-compression factor of four. 
Comparing (\ref{eq:tau_2}) and   (\ref{eq:tau_1}) we see that the expressions are very similar,
provided we interchange $V_{\rm s}$ and $V_{\rm A}$.
This belies the generally held idea that the second-order Fermi process is accelerating particles much less rapidly than DSA.
However, in astrophysical sources like young SNRs one can easily encounter $V_{\rm s}>3000$~\kms,
whereas in the Galaxy Alfv\'en velocities above \change{a few 100~\kms} are probably rare.

It is worthwhile considering why second-order Fermi acceleration is as fast as DSA for a given typical speed.
The reason is that according to DSA theory  a particle needs typically  $\sim c/V$ scatterings before it will
cross the shock front, and obtains a boost in energy. For second-order Fermi acceleration each of the scatterings, rather than  $\sim c/V$ scatterings, will  alter the energy on average by (\ref{eq:dE_E}).
Hence, the frequency of  energy boosts for second-order acceleration is $\sim c/V$ higher than for DSA,
compensating for
the smaller boost per scattering.

Fig.~\ref{fig:emax} shows the maximum energies as a function of acceleration time and as a function of $\delta$ for various values of $D_0$ and $V_{\rm A}$.
In \sect~\ref{sec:ymc} I will discuss why these values are plausible inside SBs.

Using a full CR transport equation in momentum space \citet{thornbury14} derived the approximate relation
$$
\tau_{\rm acc,2nd}\approx \frac{9}{4}\frac{D}{V_{\rm A}^2}
$$
for $\delta = 0.3$. Comparing to (\ref{eq:tau_2}) this implies that $\xi\approx 4$. 
For rigid collisions $\xi=8/3$ \citep[e.g.][]{longair11}. However, smaller values should be considered,
as one needs to question whether the ``collision time" $\Delta t$ for scatterings governing energy gains are similar to pitch angle scatterings.

\change{
Expression (\ref{eq:emax}) is purely based on  acceleration gains and does not take into account that for high enough energies the particles may escape
from the acceleration region, regarded here to be a SB. We can provide an estimate for how this limits $E_{\rm max}$ using the
criterion $\tau_{\rm acc} \approx  \tau_{\rm esc}$ \citep[sect. 3.1 in][]{hillas84}, with 
\begin{equation}\label{eq:tau_esc}
\tau_{\rm esc}= \frac{R^2}{6D},
\end{equation}
with $\tau_{\rm esc}$ the energy-dependent diffusion escape time, and $R$ the radius of the acceleration region.
Combining this with (\ref{eq:tau_2}) we find $D^2(E_{\rm max})=\delta\xi R^2V_{\rm A}^2/18$. Using the right-hand side of  (\ref{eq:diff}) we find
\begin{align}\label{eq:emax2}
E_{\rm max}=&\eta^{-1}\sqrt{\frac{1}{2}\delta\xi} ZeB R \left(\frac{V_{\rm A}}{c}\right)\\ \nonumber
=&5.5 \times 10^{14}\eta^{-1} Z \sqrt{\delta\xi}  \left(\frac{B}{10~{\rm \mu G}}\right) \left(\frac{R}{50~{\rm pc}}\right)\left(\frac{V_{\rm A}}{500~{\rm km\ s^{-1}}}\right)~{\rm eV}.
\end{align}
Since it is based on the same principles as laid out by \citet{hillas84},
this expression has the same dependence as Hillas' criterion for maximum CR energies.
It shows that for reaching a PeV in energy both $V_{\rm A}$ and $B$ needs to be relatively high. Using (\ref{eq:v_a}), we can rewrite this as
\begin{align}\label{eq:emax3}
E_{\rm max}=&\eta^{-1}\sqrt{\frac{1}{2}\delta\xi} \frac{eZB^2 R}{c \sqrt{4\pi \rho}} \\ \nonumber
=&6.4 \times 10^{14}\eta^{-1} Z\sqrt{\delta\xi}  \left(\frac{B}{10~{\rm \mu G}}\right)^2 \left(\frac{R}{50~{\rm pc}}\right)\left(
\frac{n_{\rm H}}{10^{-3}~{\rm cm^{-3}}}
\right)^{-\frac{1}{2}}~{\rm eV}.
\end{align}
Note that for $n_{\rm H}=0.001~{\rm cm^{-3}}$ and $B=10~{\rm \mu G}$, $V_{\rm A}=585$~\kms.
\chtwo{The dotted line in Fig.~\ref{fig:emax} shows the escape time for $\delta=1$ and $D=10^{26}~{\rm cm^{-2}s}$ at 100~TeV,
with the shaded area corresponding to  energies that violate the CR retention time for Bohm diffusion.
}

Simulations indicate that indeed in SBs $n_{\rm H}\approx 0.001~{\rm cm^{-3}}$ \citep[e.g.][]{oey04,krause13}.  Values of $B\gtrsim 30~{\rm \mu G}$ are needed to
reach multi-PeV energies. So under an optimistic scenario, which includes Bohm diffusion ($\eta\sim 1$) multi-PeV energies could be reached by second-order Fermi acceleration
in SBs. This is more optimistic than envisaged by \citet{vieu22} based on their stochastic acceleration scenario. Part of the reason is that they assumed a lower
magnetic-field strength and hydrodynamic turbulence with a maximum associated velocity of 100~\kms, whereas it is argued here that  $V_{\rm A}\sim 600$~\kms\ is feasible.

}


\section{The case for rapid second-order Fermi acceleration in superbubbles}
\label{sec:ymc}

\begin{figure}
  \centerline{
    \includegraphics[width=0.8\columnwidth]{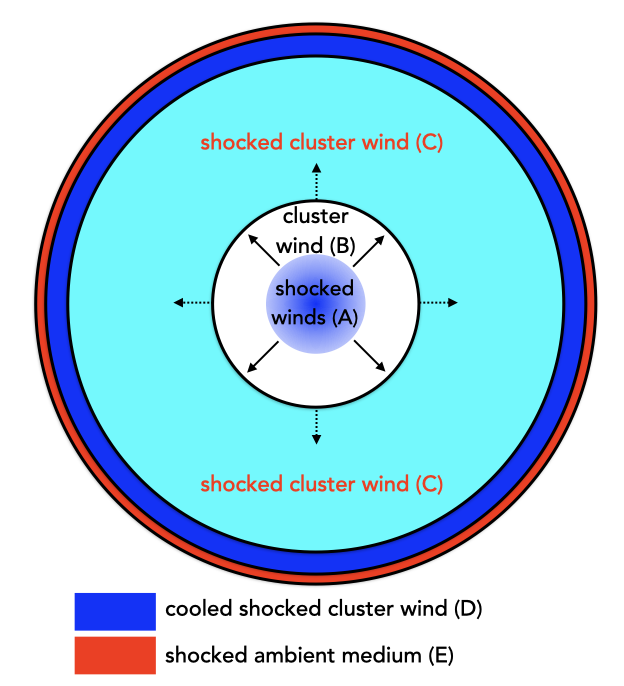}}
  \caption{\label{fig:sketch}
    Sketch of a superbubble created by a young stellar cluster \citep[after][]{koo92}.
  }
\end{figure}

\change{
CR composition studies show that for the average interstellar medium (ISM) $D_0 \approx 4\times 10^{28}$~\dunit\ for $E_0=1~{\rm GeV}/n$ \citep[e.g.][]{strong07,adriani14},
whereas the Alfv\'en speeds are at best a few tens of \kms. Indeed, \citet{thornbury14} concluded that second-order Fermi acceleration in the ISM cannot
contribute substantially to the acceleration of CRs.}
The situation in SBs is, however, very different, as I will illustrate for 
the PeVatrons HESS J1646-458/Westerlund~1 and the
Cygnus Cocoon.
\chtwo{The  importance of second-order Fermi shock acceleration  for the highest energy Galactic CRs was previously explored by \citet{bykov01b}. However, they assumed
that the acceleration was caused by multiple interacting shock and rarefaction waves, and not by high speed Alv\'en waves. }

SBs are created by young stellar clusters through the strong stellar winds of OB  and Wolf-Rayet (WR) stars, which can have a combined wind power of up to $L_{\rm w}\approx 10^{38}$--$10^{39}~{\rm erg\ s^{-1}}$.
The standard model for SBs are scaled up versions of the stellar wind bubble model by \citet{weaver77}---see \citet{chevalierclegg85,koo92}. For quick reference the basic structure is sketched
in Fig.~\ref{fig:sketch}, which shows four-five distinct regions: A) the stellar cluster itself, in which colliding stellar winds create a hot X-ray emitting plasma; B) the plasma expands into a tenuous collective 
stellar cluster wind, which ends in a termination shock; C) the shocked-heated wind creates a pressurized expanding, tenuous plasma, with densities as low as $n_{\rm H}\approx 10^{-2}$--$10^{-3}$~\pcc\
\citep{oey04,krause13}; D) and E) the SB has swept up a dense shell of ambient gas, which may be further compressed and fragment due to radiative cooling.
\chtwo{Due its high density, region D/E is the most likely emission region for gamma-ray emission, in case the radiation has a hadronic origin.}

Particle acceleration is likely occurring in various locations in the SB: DSA will occur in region A) due to colliding stellar winds and at the termination shock (the boundary of  B and C),
whereas SNR shocks will accelerated particles throughout the entire SB, as the shocks remain fast for a few thousand years. An example of the latter
may be provided by the 30~Dor~C SB, which emits nonthermal X-ray and gamma-ray radiation \citep{hess_lmc,kavanagh19}. The combination of all these regions may also re-accelerate CRs,
as they  may encounter multiple shocks over the lifetime of the SB \citep{bykov01}.

The most favorable location for second-order Fermi shock acceleration is region C, which occupies the largest  volume of the SB, and, due to its low density, 
is characterized by a relatively large Alfv\'en velocity of
the order of $V_{\rm A}=585 (B/10~{\rm \mu G})(n_{\rm H}/0.001~{\rm cm^{-3}})^{-1/2}$~\kms.
A magnetic-field strength of $B\approx 10~{\rm \mu G}$ is not unreasonable, giving measurements of $B\approx 3$--$10~{\rm \mu G}$ in the Orion-Eridanus SB \citep{joubaud19},
and the estimates of $B\approx 10$--$20~{\rm \mu G}$ in the 30~Dor~C SB \citep{kavanagh19}.
The seed particles for the second-order Fermi  process is easily provided by all the shock acceleration sites discussed above.

\begin{figure}
  \centerline{
    \includegraphics[trim=0 0 110 0,clip=true,width=0.8\columnwidth]{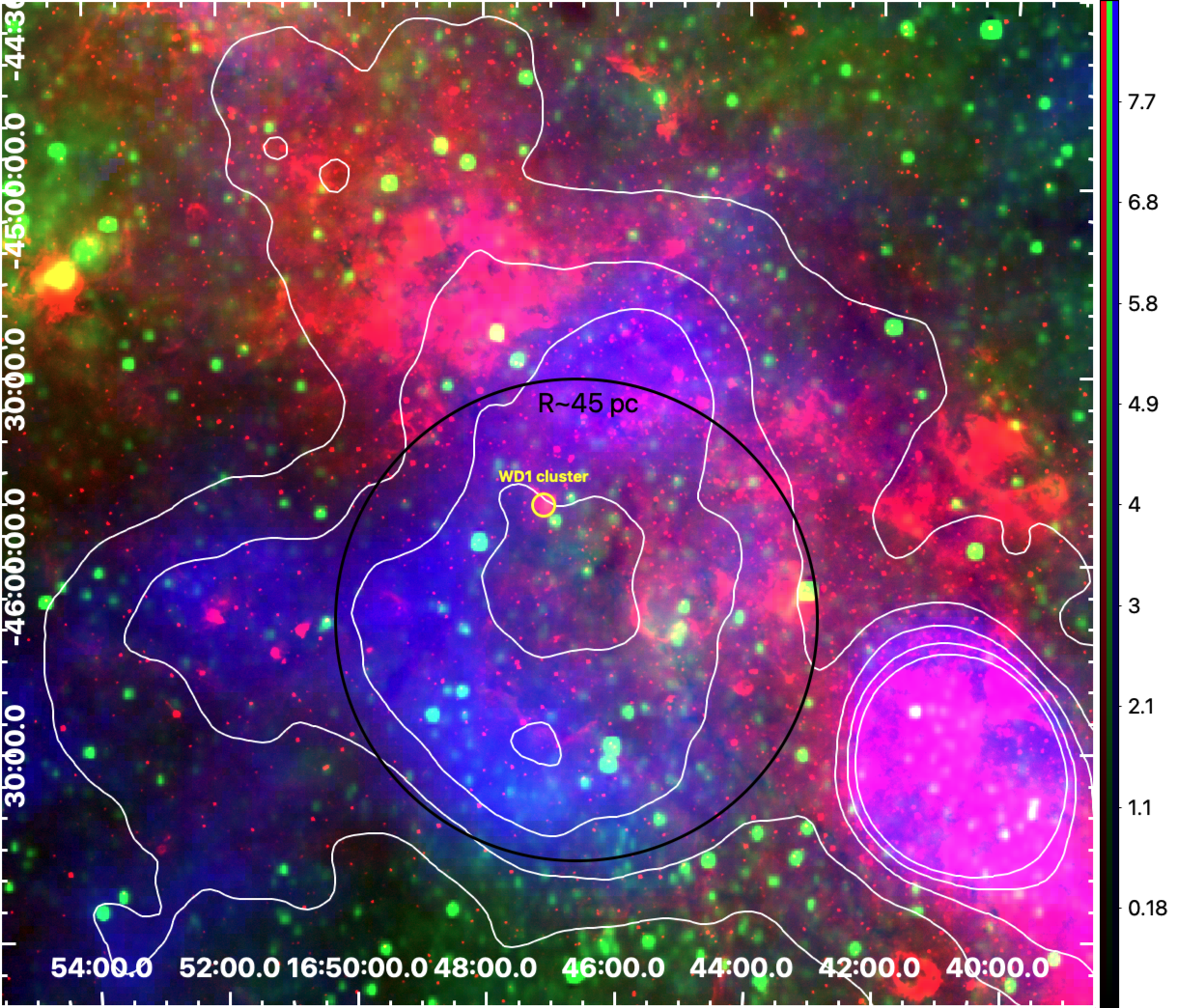}
    }
  \caption{\label{fig:wd1}
    The region around HESS J1646-458/ Westerlund 1, as seen in infrared
    \citep[red, Midcourse Space Experiment (MSX), 8.3${\mu m}$,][]{msx01},
    H$\alpha$ \citep[green, SHASSA,][]{shassa01}, and H.E.S.S. \citep[blue, based on the $>1$~TeV flux map,][]{hess22_wd1}.
    The contours show the H.E.S.S. significance contours at 5, 8 and 12 $\sigma$.
    The location of the stellar cluster Westerlund 1 itself is a bright infrared source and is indicated by a yellow circle.
The typical extent of the gamma-ray source is indicated by a black circle.
}
\end{figure}

Westerlund 1 contains 24 WR stars \citep{crowther06}. But with an age of 4~Myr, with possible longer starformation history going
back to 10~Myr \citep{navarate22}, Westerlund 1 is likely to also have been energized by supernova explosions. Cyg OB2 has three identified WR stars, and most stars seemed to have been born 2--3 Myr ago,
but starformation may have occurred over a more extended period of 7~Myr \citep{wright15}, implying additional energy input from supernovae.
The TeV gamma-ray emission from both HESS J1646-458/Westerlund~1 and the Cygnus Cocoon has a typical radius of $\approx 50$~pc. For HESS J1646-458 the morphology is shell-like,
which could either be caused by a density enhancement near the edge of SB (region D/E), 
\chtwo{implying an hadronic origin for the gamma-ray emission}, or---as favored  by \citet{hess22_wd1,haerer23}---could be identified as the termination shock 
region in case it is actively accelerating electrons up to, or beyond \chtwo{$\sim 200$~TeV}.
\chtwo{The latter interpretation  implies leptonic gamma-ray emission, as the density near the termination shock is relatively low \citep{haerer23}.
An association of the gamma-ray emission with the termination shock  also implies that the entire SB is  larger than the extent of HESS J1646-458, 
as then the  shell is associated with the B/D boundary, placing regions C, D and E much further out.}

\chtwo{In contrast,}
I express here the view that the SB is not much larger than the extent of HESS J1646-458 for two reasons: 1) HESS J1646-458 fits almost entirely within a gap
in the dense ISM, which is traced by both CO emission \citep[see Fig.~8 in][]{hess22_wd1} and 8.3~micron infrared emission, see Fig.~\ref{fig:wd1}; 2) Westerlund 1 is located northeast of the center
of HESS J1646-458, If the TeV gamma-ray emission is directly related to the termination shock one expects the brightest gamma-ray emission in the northeast,  as there the ram pressure and density of the cluster wind 
is strongest. Instead the brightest gamma-ray emission is associated with the southern part of the shell, and  in fact coincides with a bright infrared filament, which could be part of
the SB shell (Fig.~\ref{fig:wd1}). \change{Here a hadronic origin for the gamma-ray emission is favored, as it more naturally explains the uniformity of the gamma-ray spectrum throughout the source
\citet{hess22_wd1}. See, however, \citet{haerer23} for arguments for a leptonic origin of the gamma-ray emission. For the Cygnus Cocoon the hadronic nature of the gamma-ray emission
is not disputed, and  recently \citet{vieu24} showed through hydrodynamical simulations that the stellar cluster is too extended for a powerful, collective
cluster wind.
}

Irrespective of the interpretation of the shell-like morphology, the typical radius of $R\approx 45$~pc for a distance of 4~kpc \citep{navarate22} implies a strongly suppressed value of the diffusion coefficient.
The energy in CR protons is estimated to be $W_{\rm p}\approx 6\times 10^{51} n_{\rm H}^{-1}$~erg \chtwo{for $E_{\rm cr}>1$~GeV and a particle power-law index of $\Gamma=2.33$} \citep{hess22_wd1}, which should be compared to an expected overall energy input from stellar winds of
$E_{\rm w}\approx L_{\rm w}t\approx 3\times 10^{52} (L_{\rm w}/10^{39}~{\rm erg\ s^{-1}})t_6$~erg, with $L_{\rm w}\approx 10^{39}~{\rm erg\ s^{-1}}$, 
the current wind luminosity of Westerlund 1 \citep{muno06}, \change{and $t_6$ the age of the SB in Myr.} 
Not only do these values imply a CR efficiency of 5--20\%, it also means that the CRs \change{responsible for the TeV gamma-ray emission} must have been
largely confined to the gamma-ray emitting region over the lifetime of the SB. This provides another reason why the size of the SB cannot exceed the gamma-ray emitting region by a large fraction.
If the SB was formed over the last 1~Myr of the stellar cluster, we can use (\ref{eq:tau_esc}) to estimate the diffusion coefficient to be
\change{$D \approx  1.0 \times 10^{26} (R/45~{\rm pc})^2  t_6^{-1}~{\rm cm^2\ s^{-1}}$ for typical proton energies of 100~TeV.} 
\change{For $\delta=0.5$ the ISM value of $D_0$ extrapolated to 100~TeV implies $D({\rm 100~TeV})\approx 10^{31}$~\dunit, which is five orders of magnitude larger than for
 HESS J1646-458.}
As already noted in \citet{hess22_wd1} the small diffusion coefficient implies Bohm diffusion \change{and a relatively strong magnetic field strength:
for the inferred $D(100~{\rm TeV})\approx 10^{21}~{\rm cm^{-3}s}$, Eq. (\ref{eq:diff}) implies $B\gtrsim 30~{\rm \mu G}$. }

Note that  a highly turbulent magnetic-field strength of $B=30~{\rm \mu G}$ 
 has an associated energy of 
$E_B\approx 
\frac{1}{6}R^3(\delta B)^2 \approx 4\times 10^{50}(\delta B/30~{\rm  \mu G})^2(R/45~{\rm pc})^3$~erg, which is  $\sim 1$\%  of $E_{\rm w}$.
\chtwo{On the other hand, this energy is  an order of magnitude smaller than the  inferred CR energy of HESS J1646-458 \citep{hess22_wd1}. 
This does not  violate the energetic constraints of
second-order Fermi acceleration: the CR energy is dominated by low-energy CRs ($\lesssim 100$~GeV),
which likely have their origin in DSA in the cluster itself or the termination shock. 
In addition,
for second-order Fermi acceleration a continuous exchange of energy from turbulent magnetic field to CR particles is required. 
As a result, the  energy in CRs is dominated by cumulative effects, as long as CR energy escape
is limited, whereas the magnetic-field energy reflects the current status, which is governed by the dynamic interplay
between turbulent magnetic-field generation, and damping effects. The latter includes the transfer of turbulent magnetic-field energy to CR energy.
}

The  discussion on HESS J1646-458/Westerlund 1 is also applicable to the Cygnus  Cocoon. Also for the Cygnus Cocoon the emission comes from a region that has a low infrared flux, and is surrounded by
bright infrared emission, which suggests that the SB size is comparable to the gamma-ray emitting Cygnus Cocoon itself.  \citet{hawc21_cygnus} estimated the diffusion coefficient to be  much smaller
than that of the ISM. Using $D=R^2/6t$ and $R=55$~pc, $t=1$~Myr yields $D=1.5\times 10^{26}$~\dunit. \change{Also in this case a relative large value for the magnetic-field strength, $B\gtrsim 20~{\rm \mu G}$, is required in order to meet the Bohm diffusion criterion.  A somewhat larger, but still small diffusion coefficient was inferred by \citet{lhaaso24_cygnus}, who estimated $D\approx 3\times 10^{26}$~\dunit
  for an energy of 1~TeV.}

The clear indications for small diffusion coefficients coupled to the need for a relatively strong magnetic field of $B>10~{\rm \mu G}$ shows that \change{SBs in general may} provide the right
conditions for efficient second-order Fermi acceleration:\change{
at 100~TeV values for the diffusion coefficients \change{ may be} $D\sim 10^{26}$--$10^{27}$~\dunit,}
whereas Alfv\'en speeds may be of the order of
of  $V_{\rm A}=370$~\kms\ for $B=20~{\rm \mu G}, n_{\rm H}=0.01$~\pcc, or even $\sim1000$~\kms\ if densities are as low as $n_{\rm H}=0.001$~\pcc.
Eq.~(\ref{eq:emax}) shows, and  Fig.~\ref{fig:emax} illustrates, that these values are sufficient to accelerate CRs \change{from 1~TeV to  a few PeV.} 
The limiting factor is  the size of the SB region, and the strength and turbulence of the turbulent magnetic-field, which govern the diffusive escape
of CRs. 
\change{Fig.~\ref{fig:emax} also shows that due escape, the highest values for $E_{\rm max}$ need to be reached within relatively short time scales. The diffusive escape
also requires a spectral steepening toward the highest energies. This, indeed, seems to be the case for the Cygnus Cocoon \citep{lhaaso24_cygnus}.}


\section{Conclusions}

I have argued here that second-order Fermi acceleration can be a rapid acceleration mechanism, provided that the Alv\'en speed is high ($V_{\rm A}\gtrsim 300$~\kms) and that the diffusion coefficient is small
\change{($D\approx 10^{26}$~\dunit at 100 TeV).}
Second-order Fermi acceleration is often regarded as a slow acceleration mechanism. But if one compares the relevant  speeds, $V_{\rm A}$ versus shock speed $V_{\rm s}$,
the mechanism can be as rapid as first order acceleration. 

In practice the Universe is more likely to provide sites with $V_{\rm s}\gtrsim 1000$~\kms\ then  regions with \change{
$V_{\rm A}\gtrsim$ a few $100$~\kms. } \change{For the ISM $V_{\rm A}\sim  10$~\kms,  which is unfavorable for second-order Fermi acceleration.}
 However, the conditions are very different in SBs, given the estimated the Alv\'en speed and diffusion coefficients in two Galactic 
PeVatron sources,  HESS J1646-458/Westerlund 1 and 
the Cygnus Cocoon/Cyg OB2. This indicates that second-order Fermi acceleration can provide the mechanism to turn SBs into PeVatrons.
As such it provides an alternative, or complementary theory, \change{ to the  association of PeVatrons with DSA by the cluster-wind termination shock \citep{morlino21,vieu22,blasi23,haerer23},
or SNRs inside SBs \citep{vieu23}.}

\change{
Note that  I concentrated on the question whether second-order Fermi acceleration can accelerate CRs up to or beyond the CR ``knee", but did not
elaborate on the expected CR spectrum that may result from it. Unlike DSA, second-order Fermi acceleration does not have a, scale-free, built-in escape mechanism---for DSA this is due to downstream advection.
Intrinsically, second-order Fermi acceleration
therefore, results in hard CR spectra. However, there is diffusive, energy-dependent escape, which results in a gradual steepening of the spectrum. \chtwo{Moreover, the detailed
starformation history of the SB and the shape of the injection spectrum also play a role in  shaping the very-high-energy CR spectrum.}
The combined effect of energy dependent escape and the history of CR energy input is complex and requires further investigation.}

Finally, it is worth pointing out that besides SBs there may be other locations where the right level of turbulence, and high Alfv\'en velocities exist that are favorable for  second-order Fermi
acceleration. \change{For example, the magnetic-field strengths in the Galactic Center is higher than 50~${\rm \mu G}$ \citep{crocker10}, 
and for starbursts galaxies mG magnetic-field strengths
have been inferred \citep{mcbride14}. If these values also apply to their low-density ISM, Alfv\'en speeds may be high, and coupled to turbulent magnetic fields, the Galactic Center and starburst galaxies
may also contain  locations for CR acceleration beyond PeV energies.}


\begin{acknowledgements}
I  appreciate helpful discussion during the Vulcano Workshop 2022 at Elba, and at the workshop ``Supernova Remnants in Complex Environments" at the Lorentz Center Leiden (October 2023).
For this paper I made use of the Midcourse Space Experiment (MSX) archive
(\url{https://irsa.ipac.caltech.edu/applications/MSX/MSX/}), the
H.E.S.S. map of Westerlund 1 in FITS (\url{https://www.mpi-hd.mpg.de/hfm/HESS/pages/publications/auxiliary/2022_Westerlund1}), and the
  SkyView is a Virtual Observatory (\url{https://skyview.gsfc.nasa.gov}).
This research has made use of the NASA/IPAC Infrared Science Archive, which is operated by the Jet Propulsion Laboratory, California Institute of Technology, under contract with NASA.

\end{acknowledgements}



\begin{thebibliography}{47}
\expandafter\ifx\csname natexlab\endcsname\relax\def\natexlab#1{#1}\fi

\bibitem[{{Abdalla} {et~al.}(2021)}]{hess21_1702}
{Abdalla}, H. {et~al.} 2021, \aap, 653, A152

\bibitem[{{Abeysekara} {et~al.}(2021){Abeysekara}, {Albert}, {Alfaro},
  {Alvarez}, {Camacho}, {Arteaga-Vel{\'a}zquez}, {et~al.}}]{hawc21_cygnus}
{Abeysekara}, A.~U., {Albert}, A., {Alfaro}, R., {et~al.} 2021, Nature
  Astronomy, 5, 465

\bibitem[{{Abramowski} {et~al.}(2015){Abramowski}, {Aharonian}, {Ait Benkhali},
  {Akhperjanian}, \& et~al.}]{hess_lmc}
{Abramowski}, A., {Aharonian}, F., {Ait Benkhali}, F., {Akhperjanian}, A.~G.,
  \& et~al. 2015, Science, 347, 406

\bibitem[{{Ackermann} {et~al.}(2011){Ackermann}, {Ajello}, {Allafort},
  {Baldini}, {Ballet}, {et~al.}}]{fermi11_cocoon}
{Ackermann}, M., {Ajello}, M., {Allafort}, A., {et~al.} 2011, Science, 334,
  1103

\bibitem[{{Adriani} {et~al.}(2014){Adriani}, {Barbarino}, {Bazilevskaya},
  {Bellotti}, {Boezio}, {et~al.}}]{adriani14}
{Adriani}, O., {Barbarino}, G.~C., {Bazilevskaya}, G.~A., {et~al.} 2014, \apj,
  791, 93

\bibitem[{{Aharonian} {et~al.}(2022)}]{hess22_wd1}
{Aharonian}, F. {et~al.} 2022, \aap, 666, A124

\bibitem[{{Albert} {et~al.}(2020){Albert}, {Alfaro}, {Alvarez}, {Camacho},
  {Arteaga-Vel{\'a}zquez}, {et~al.}}]{hawc20}
{Albert}, A., {Alfaro}, R., {Alvarez}, C., {et~al.} 2020, \apjl, 896, L29

\bibitem[{{Blasi} \& {Morlino}(2023)}]{blasi23}
{Blasi}, P. \& {Morlino}, G. 2023, \mnras, 523, 4015

\bibitem[{{Bykov}(2001)}]{bykov01}
{Bykov}, A.~M. 2001, \ssr, 99, 317

\bibitem[{{Bykov} \& {Fleishman}(1992)}]{bykov92}
{Bykov}, A.~M. \& {Fleishman}, G.~D. 1992, \mnras, 255, 269

\bibitem[{{Bykov} \& {Toptygin}(2001)}]{bykov01b}
{Bykov}, A.~M. \& {Toptygin}, I.~N. 2001, Astronomy Letters, 27, 625

\bibitem[{{Cao} {et~al.}(2021)}]{lhaaso21}
{Cao}, Z., A. F. A.~Q. {et~al.} 2021, Nature, 594

\bibitem[{{Cao} {et~al.}(2024){Cao}, {Aharonian}, {An}, {Axikegu}, {Bai},
  {et~al.}}]{lhaaso24}
{Cao}, Z., {Aharonian}, F., {An}, Q., {et~al.} 2024, \apjs, 271, 25

\bibitem[{{Chevalier} \& {Clegg}(1985)}]{chevalierclegg85}
{Chevalier}, R.~A. \& {Clegg}, A.~W. 1985, \nat, 317, 44

\bibitem[{{Cristofari}(2021)}]{cristofari21}
{Cristofari}, P. 2021, Universe, 7, 324

\bibitem[{{Crocker} {et~al.}(2010){Crocker}, {Jones}, {Melia}, {Ott}, \&
  {Protheroe}}]{crocker10}
{Crocker}, R.~M., {Jones}, D.~I., {Melia}, F., {Ott}, J., \& {Protheroe}, R.~J.
  2010, \nat, 463, 65

\bibitem[{{Crowther} {et~al.}(2006){Crowther}, {Hadfield}, {Clark},
  {Negueruela}, \& {Vacca}}]{crowther06}
{Crowther}, P.~A., {Hadfield}, L.~J., {Clark}, J.~S., {Negueruela}, I., \&
  {Vacca}, W.~D. 2006, \mnras, 372, 1407

\bibitem[{{Fermi}(1949)}]{fermi49}
{Fermi}, E. 1949, Physical Review, 75, 1169

\bibitem[{{Gabici} \& {Aharonian}(2007)}]{gabici07}
{Gabici}, S. \& {Aharonian}, F.~A. 2007, \apjl, 665, L131

\bibitem[{{Gaustad} {et~al.}(2001){Gaustad}, {McCullough}, {Rosing}, \& {Van
  Buren}}]{shassa01}
{Gaustad}, J.~E., {McCullough}, P.~R., {Rosing}, W., \& {Van Buren}, D. 2001,
  \pasp, 113, 1326

\bibitem[{{H{\"a}rer} {et~al.}(2023){H{\"a}rer}, {Reville}, {Hinton},
  {Mohrmann}, \& {Vieu}}]{haerer23}
{H{\"a}rer}, L.~K., {Reville}, B., {Hinton}, J., {Mohrmann}, L., \& {Vieu}, T.
  2023, \aap, 671, A4

\bibitem[{{Hawc} {et~al.}(2022){Hawc}, {Abeysekara}, {Albert}, {Alfaro},
  {Alvarez}, {et~al.}}]{hawc22_gro}
{Hawc}, {Abeysekara}, A.~U., {Albert}, A., {et~al.} 2022, in 37th International
  Cosmic Ray Conference, 810

\bibitem[{{Helder} {et~al.}(2012){Helder}, {Vink}, {Bykov}, {Ohira}, {Raymond},
  \& {Terrier}}]{helder12a}
{Helder}, E.~A., {Vink}, J., {Bykov}, A.~M., {et~al.} 2012, \ssr, 173, 369

\bibitem[{{Hillas}(1984)}]{hillas84}
{Hillas}, A.~M. 1984, \araa, 22, 425

\bibitem[{{Joubaud} {et~al.}(2019){Joubaud}, {Grenier}, {Ballet}, \&
  {Soler}}]{joubaud19}
{Joubaud}, T., {Grenier}, I.~A., {Ballet}, J., \& {Soler}, J.~D. 2019, \aap,
  631, A52

\bibitem[{{Kavanagh} {et~al.}(2019){Kavanagh}, {Vink}, {Sasaki}, {Chu},
  {Filipovi{\'c}}, {Ohm}, {Haberl}, {Manojlovic}, \& {Maggi}}]{kavanagh19}
{Kavanagh}, P.~J., {Vink}, J., {Sasaki}, M., {et~al.} 2019, \aap, 621, A138

\bibitem[{{Koo} \& {McKee}(1992)}]{koo92}
{Koo}, B.-C. \& {McKee}, C.~F. 1992, \apj, 388, 93

\bibitem[{{Krause} {et~al.}(2013){Krause}, {Fierlinger}, {Diehl}, {Burkert},
  {Voss}, \& {Ziegler}}]{krause13}
{Krause}, M., {Fierlinger}, K., {Diehl}, R., {et~al.} 2013, \aap, 550, A49

\bibitem[{{Lhaaso Collaboration}(2024)}]{lhaaso24_cygnus}
{Lhaaso Collaboration}. 2024, Science Bulletin, 69, 449

\bibitem[{{Longair}(2011)}]{longair11}
{Longair}, M.~S. 2011, {High Energy Astrophysics} (Cambridge University Press,
  2011)

\bibitem[{{MAGIC Collaboration} {et~al.}(2023){MAGIC Collaboration}, {Abe},
  {Abe}, {Acciari}, {Agudo}, {Aniello}, {et~al.}}]{magic21_g109}
{MAGIC Collaboration}, {Abe}, H., {Abe}, S., {et~al.} 2023, \aap, 671, A12

\bibitem[{{McBride} {et~al.}(2014){McBride}, {Quataert}, {Heiles}, \&
  {Bauermeister}}]{mcbride14}
{McBride}, J., {Quataert}, E., {Heiles}, C., \& {Bauermeister}, A. 2014, \apj,
  780, 182

\bibitem[{{Montmerle}(1979)}]{montmerle79}
{Montmerle}, T. 1979, \apj, 231, 95

\bibitem[{{Morlino} {et~al.}(2021){Morlino}, {Blasi}, {Peretti}, \&
  {Cristofari}}]{morlino21}
{Morlino}, G., {Blasi}, P., {Peretti}, E., \& {Cristofari}, P. 2021, \mnras,
  504, 6096

\bibitem[{{Muno} {et~al.}(2006){Muno}, {Law}, {Clark}, {Dougherty}, {de Grijs},
  {Portegies Zwart}, \& {Yusef-Zadeh}}]{muno06}
{Muno}, M.~P., {Law}, C., {Clark}, J.~S., {et~al.} 2006, \apj, 650, 203

\bibitem[{{Navarete} {et~al.}(2022){Navarete}, {Damineli}, {Ramirez}, {Rocha},
  \& {Almeida}}]{navarate22}
{Navarete}, F., {Damineli}, A., {Ramirez}, A.~E., {Rocha}, D.~F., \& {Almeida},
  L.~A. 2022, \mnras, 516, 1289

\bibitem[{{Oey} \& {Garc{\'\i}a-Segura}(2004)}]{oey04}
{Oey}, M.~S. \& {Garc{\'\i}a-Segura}, G. 2004, \apj, 613, 302

\bibitem[{{Parizot} {et~al.}(2004){Parizot}, {Marcowith}, {van der Swaluw},
  {Bykov}, \& {Tatischeff}}]{parizot04}
{Parizot}, E., {Marcowith}, A., {van der Swaluw}, E., {Bykov}, A.~M., \&
  {Tatischeff}, V. 2004, \aap, 424, 747

\bibitem[{{Price} {et~al.}(2001){Price}, {Egan}, {Carey}, {Mizuno}, \&
  {Kuchar}}]{msx01}
{Price}, S.~D., {Egan}, M.~P., {Carey}, S.~J., {Mizuno}, D.~R., \& {Kuchar},
  T.~A. 2001, \aj, 121, 2819

\bibitem[{{Strong} {et~al.}(2007){Strong}, {Moskalenko}, \&
  {Ptuskin}}]{strong07}
{Strong}, A.~W., {Moskalenko}, I.~V., \& {Ptuskin}, V.~S. 2007, Annual Review
  of Nuclear and Particle Science, 57, 285

\bibitem[{{Thornbury} \& {Drury}(2014)}]{thornbury14}
{Thornbury}, A. \& {Drury}, L.~O. 2014, \mnras, 442, 3010

\bibitem[{{Vieu} {et~al.}(2024){Vieu}, {Larkin}, {H{\"a}rer}, {Reville},
  {Sander}, \& {Ramachandran}}]{vieu24}
{Vieu}, T., {Larkin}, C. J.~K., {H{\"a}rer}, L., {et~al.} 2024, arXiv e-prints,
  arXiv:2406.13589

\bibitem[{{Vieu} \& {Reville}(2023)}]{vieu23}
{Vieu}, T. \& {Reville}, B. 2023, \mnras, 519, 136

\bibitem[{{Vieu} {et~al.}(2022){Vieu}, {Reville}, \& {Aharonian}}]{vieu22}
{Vieu}, T., {Reville}, B., \& {Aharonian}, F. 2022, \mnras, 515, 2256

\bibitem[{Vink(2020)}]{vinkbook}
Vink, J. 2020, Physics and Evolution of Supernova Remnants, Astronomy and
  Astrophysics Library (Springer International Publishing)

\bibitem[{{Weaver} {et~al.}(1977){Weaver}, {McCray}, {Castor}, {Shapiro}, \&
  {Moore}}]{weaver77}
{Weaver}, R., {McCray}, R., {Castor}, J., {Shapiro}, P., \& {Moore}, R. 1977,
  \apj, 218, 377

\bibitem[{{Wright} {et~al.}(2015){Wright}, {Drew}, \& {Mohr-Smith}}]{wright15}
{Wright}, N.~J., {Drew}, J.~E., \& {Mohr-Smith}, M. 2015, \mnras, 449, 741

\end{thebibliography}

\end{document}